\input{aipcheck}

\documentclass[
    ,final            
  ]
  {aipproc}

\layoutstyle{8x11double}

\begin{document}

\title{Non-linear viscous saturation of r-modes}


\author{Mark Alford}{
  address={Department of Physics, Washington University in St. Louis, Missouri, 63130, USA}
}

\author{Simin Mahmoodifar}{
  address={Department of Physics, Washington University in St. Louis, Missouri, 63130, USA}
}

\author{Kai Schwenzer}{
  address={Department of Physics, Washington University in St. Louis, Missouri, 63130, USA}
}

\begin{abstract}
Pulsar spin frequencies and their time evolution are an important source of information on compact stars and their internal composition. Oscillations of the star can reduce the rotational energy via the emission of gravitational waves. 
In particular unstable oscillation modes, like r-modes, are relevant since their amplitude becomes large and can lead to a fast spin-down of young stars if they are saturated by a non-linear saturation mechanism. We present a novel mechanism based on the pronounced large-amplitude enhancement of the bulk viscosity of dense matter. We show that the enhanced damping due to non-linear bulk viscosity can saturate r-modes of neutron stars at amplitudes appropriate for an efficient spin-down.
\end{abstract}

\maketitle


Compact stars are so dense that they could contain deconfined quark matter. To reveal their decomposition requires to connect microscopic properties to macroscopic observables. R-modes are oscillations modes of rotating stars that are unstable under the emission of gravitational waves \cite{Papaloizou:1978zz,Andersson:1997xt,Andersson:2000mf,Lindblom:1999yk}. In case of neutron stars they could spin-down young stars within a short time interval or limit the rotational rate of old stars that are spun up by accretion. However, since viscous damping cannot stop low amplitude r-modes in certain instability regions at high frequency they have to be saturated by some non-linear damping mechanism. We propose the strong suprathermal enhancement of the bulk viscosity at large amplitudes \cite{Madsen:1992sx,Alford:2010gw} as a viable saturation mechanism and in contrast to previous studies \cite{Reisenegger:2003pd} find that it can stop the growth of r-modes of neutron stars at amplitudes that are large enough for an efficient spin-down but small enough that viscous damping could dominate competing saturation mechanisms \cite{Bondarescu:2008qx,Lin:2004wx}.

The bulk viscosity of dense matter provides a measure for the energy dissipation in a compression and rarefaction cycle. It is maximal when the external oscillation frequency matches the time scale of the microscopic interactions that cause the dissipation, whereby the dominant interactions in the case of star oscillations are slow weak processes. The external density fluctuation $\Delta n$ induces a corresponding oscillation of the difference of chemical potentials $\mu_\Delta$ which would vanish in weak equilibrium. Similarly, the rate of weak interactions vanishes generally for fully degenerate matter in equilibrium and becomes finite either due to finite temperature effects or due to such deviations of the chemical potentials from their equilibrium value. The net rate for the considered weak process $\Gamma^{(\leftrightarrow)}$ takes the general form
\begin{equation}
\Gamma^{(\leftrightarrow)} = -\tilde{\Gamma} T^{2N}\mu_\Delta \left(1+\sum_{j=1}^{N}\chi_j\left(\frac{\mu_\Delta^{2}}{T^{2}}\right)^{j}\right)\,.
\label{eq:gamma-parametrization}\end{equation}
where the coefficients $\chi_j$ of the terms that are non-linear in the oscillating chemical potential difference $\mu_\Delta$ and their number $N$ are determined by the respective weak process. In this work we study neutron matter with an APR equation of state \cite{Akmal:1998cf} at densities that are not high enough to allow direct Urca reactions, so that only the modified Urca process involving a bystander nucleon $n+n \rightarrow n+p+e^- +\bar\nu_e\, , \,\cdots $ is kinematically allowed. For this process $N=3$, corresponding to a strong non-linear dependence on $T$ and $\mu_\Delta=\mu_n-\mu_p-\mu_e$. 

The bulk viscosity features three distinct characteristic regions \cite{Madsen:1992sx,Alford:2010gw}. In the {\em subthermal} regime at low amplitudes, where $\mu_\Delta \ll T$, $\mu_\Delta$ is linear in $\Delta n$ and the viscosity is independent of the amplitude. The {\em suprathermal} regime $\mu_\Delta>T$ is divided into a part where $\mu_\Delta$ is still linear in $\Delta n$ but the viscosity strongly rises and another where the rise of $\mu_\Delta$ becomes weaker due to non-linear saturation effects and the viscosity decreases again. Previously mainly the subthermal limit $\mu_\Delta \ll T$ has been studied, but since the r-mode rises exponentially it eventually reaches the suprathermal regime that will be studied in the following.

\begin{figure}
\begin{minipage}[t]{0.52\textwidth}%
\includegraphics[scale=0.96]{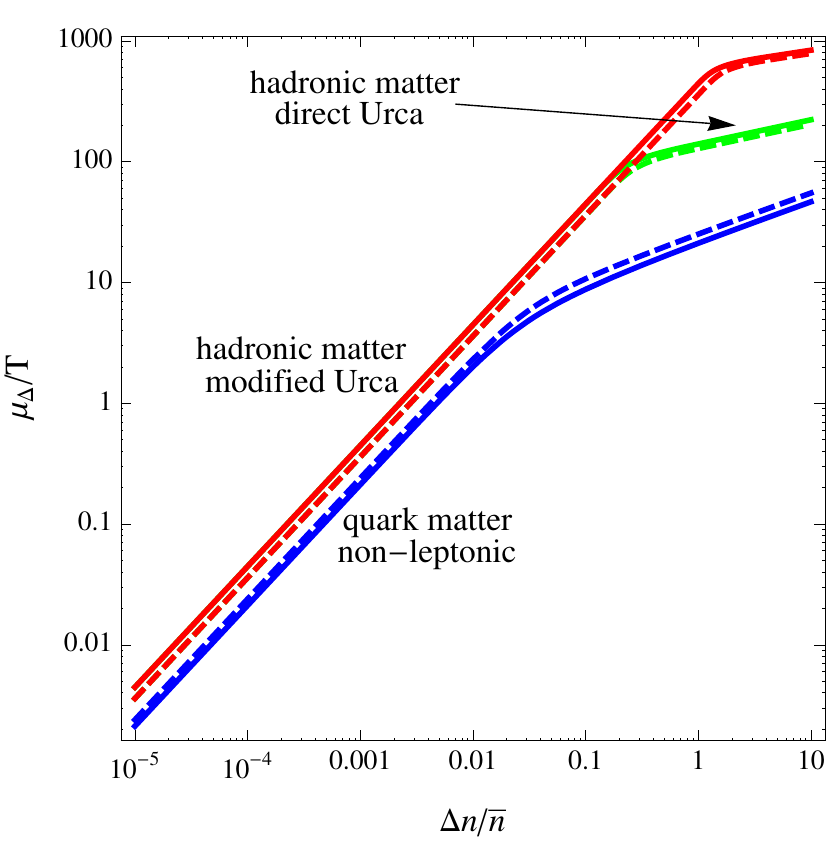}%
\end{minipage}%
\begin{minipage}[t]{0.52\textwidth}%
\includegraphics[scale=1.]{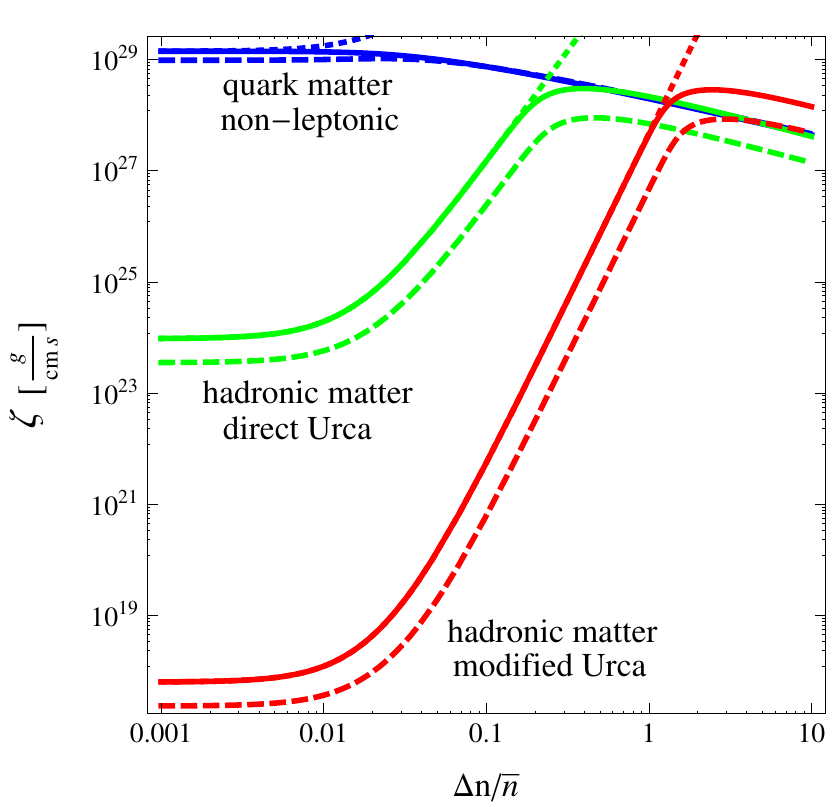}%
\end{minipage}
\caption{\label{fig:all-viscosity}
{\em Left panel:} The amplitude of the chemical potential oscillation $\mu_\Delta/T$ as a function of the driving density oscillation amplitude $\Delta n/\bar{n}$ for different forms of dense matter. {\em Right panel:} Comparison of the bulk viscosity of the different forms of matter as a function of the
density oscillation amplitude. In both plots, the frequency is
$\omega=8.4$~kHz, corresponding to an r-mode in a millisecond pulsar, and characteristic values $T=10^9$ K of the temperature and $\bar n = 2n_0$ for the density are chosen. The solid curves show the result for interacting matter, the dashed curved for non-interacting matter, whereas the dotted lines denote the analytic approximation eq. (\ref{eq:madsen-approximation}); for more details see \cite{Alford:2010gw}.}
\end{figure}

At sufficiently low temperature and large frequency the bulk viscosity allows a general analytic approximation that covers both the subthermal and the part of the suprathermal regime where $\mu_\Delta$ is linear in $\Delta n/\bar{n}$ \cite{Madsen:1992sx,Alford:2010gw,Reisenegger:2003pd}
\begin{equation}
\label{eq:madsen-approximation}
\zeta^\sim = \frac{C^{2}\tilde{\Gamma}T^{2N}}{\omega^{2}}\left(1+\sum_{j=1}^{N}\frac{\left(2j+1\right)!! \chi_{j}}{2^{j}\left(j+1\right)!} \left(\frac{C}{T} \frac{\Delta n}{\bar{n}}\right)^{2j}\right)
\end{equation}
where $C$ is a susceptibility describing the strongly interacting state of matter which in the case of neutron star matter takes the form
\begin{equation}
C\equiv\bar{n}\left.\frac{\partial\mu_\Delta}{\partial n}\right|_x = 4\!\left(1\!-\!2x\right)\!\left(\! n\!\frac{\partial S}{\partial n}\!-\!\frac{S}{3}\!\right)
\end{equation}
in terms of the symmetry energy $S$, the baryon density $n$ and the proton fraction $x$. 

The validity of the approximation eq. (\ref{eq:madsen-approximation}) can be seen from fig. \ref{fig:all-viscosity}, where on the left panel the amplitude of the chemical potential oscillation $\mu_\Delta/T$ is plotted against the corresponding density amplitude $\Delta n/\bar{n}$. For the different forms of dense matter shown there the suprathermal regime is reached for amplitudes $\Delta n/\bar{n}=O(0.01)$. In particular, in the case of hadronic matter with modified Urca interactions the linear regime, where $\mu_\Delta\sim\Delta n/\bar{n}$, extends at large frequencies basically over the entire physical range of density amplitudes $\Delta n/\bar{n}<1$. Correspondingly, the approximation eq. (\ref{fig:all-viscosity}) shown by the dotted curve on the right panel of fig. \ref{fig:all-viscosity} reproduces the full result shown by the solid curve favorably in this case.

\begin{figure}
\begin{minipage}[t]{0.52\textwidth}%
\includegraphics[scale=1.]{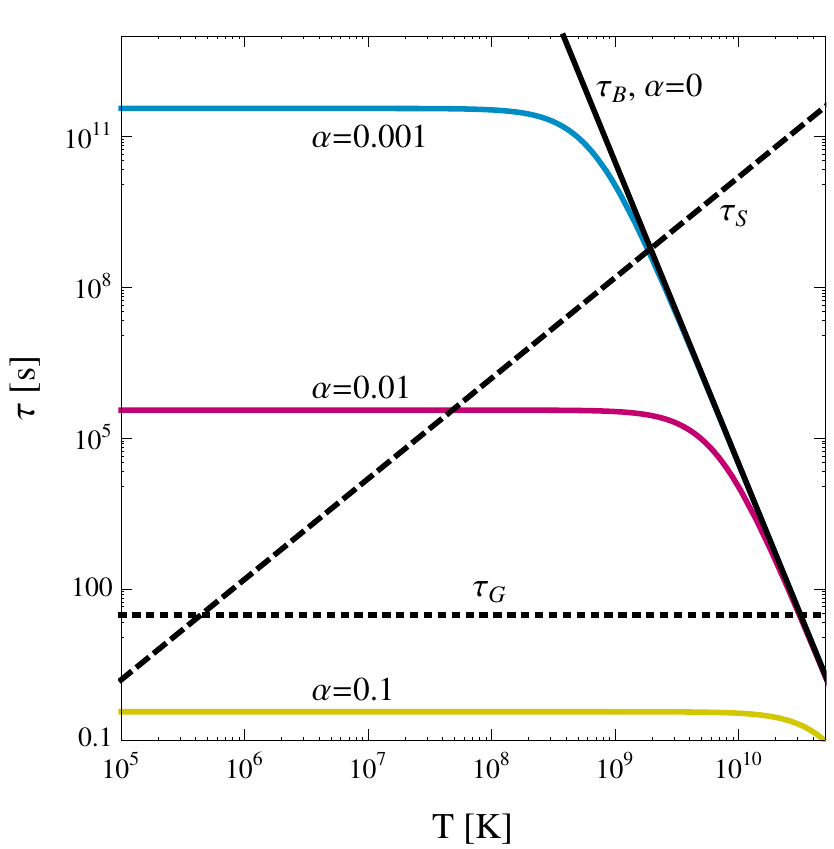}%
\end{minipage}%
\begin{minipage}[t]{0.52\textwidth}%
\includegraphics[scale=1.]{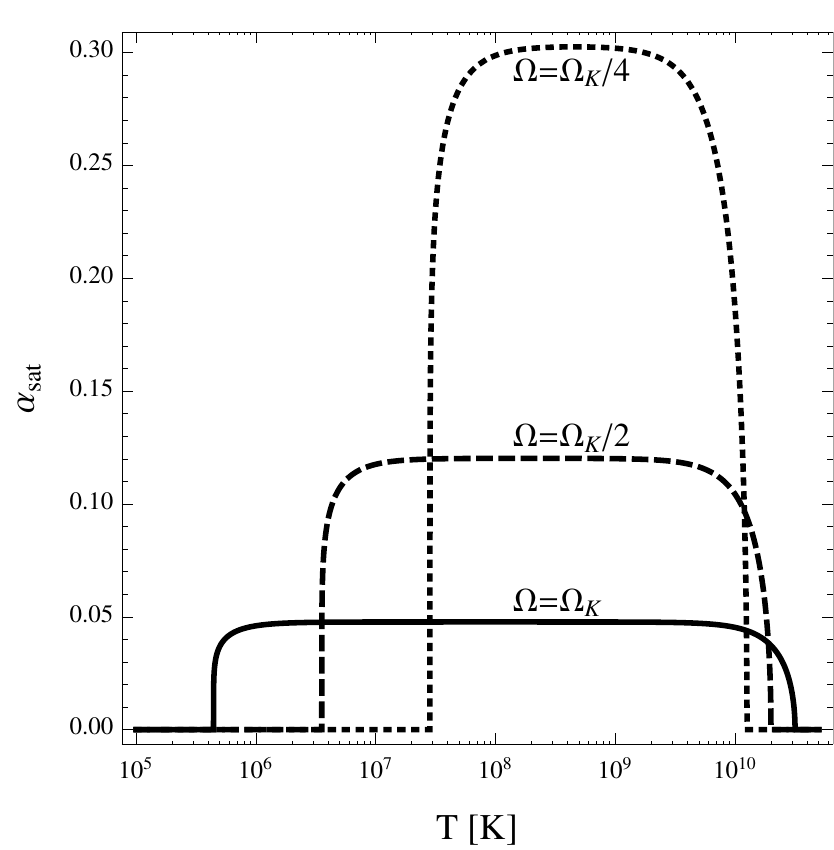}
\end{minipage}
\caption{\label{fig:saturation-amplitude} {\em Left panel:} Time scales of the fundamental $m=2$ r-mode of a  $1.4\,M_\odot$ neutron star as a function of temperature. The dotted curve represents the gravitational time scale, the dashed curve the shear viscosity damping time scale and the solid curves the bulk viscosity time scales at various amplitudes. {\em Right panel:} Temperature dependence of the "static" saturation amplitude. 
The solid, dashed, and dotted curve are given for the maximum Kepler frequency $\Omega_K\approx 6$ kHz and lower values.
}
\end{figure}

We consider in the following an exemplary  $1.4\,M_\odot$ neutron star model obtained as a solution of the TOV equations \cite{Tolman:1939jz} 
using the APR equation of state \cite{Akmal:1998cf}. The analysis of the bulk viscosity damping time requires in principle an expression for the r-mode to next to leading order in a slow rotation expansion \cite{Lindblom:1999yk}. Following \cite{Lindblom:1998wf,Jaikumar:2008kh} we restrict ourselves to the leading order expression for the fundamental $m=2$ r-mode 
with density fluctuation
\begin{equation}
\label{eq:r-mode-profile}
\left| \frac{\Delta n}{n} \right|\approx\sqrt{\frac{4\pi5!}{7}}\alpha R^{2}\Omega^{2} \frac{\partial\rho}{\partial p}\left(\left(\frac{r}{R}\right)^{3}+\delta\Phi_{0}\right)Y_{3}^{2}\left(\theta,\phi\right)
\end{equation}
where $\alpha$ is a dimensionless amplitude parameter and $\delta\Phi_0$ denotes the deviation of the gravitational potential from the equlilibrium. Eq. (\ref{eq:r-mode-profile}) has a very strong radial dependence both due to the explicit cubic factor as well due to the pronounced r-dependence of the pressure derivative of the energy density $\partial\rho/\partial p$ in a neutron star.
The characteristic time scales of the gravitational instability $\tau_G$ and of the damping by shear and bulk viscosity $\tau_S$ and $\tau_B$ are defined via their energy dissipation
\begin{equation}
\frac{1}{\tau_{i}}\equiv-\frac{1}{2E}\left(\frac{dE}{dt}\right)_{i}
\end{equation}
The stability criterium for the r-mode is $1/\tau_G+1/\tau_S+1/\tau_V\geq0$ where the equality defines the boundary of the instability region. At small amplitudes in the subthermal regime of the bulk viscosity there is at high frequency an instability region where the r-mode is unstable and grows exponentially \cite{Andersson:2000mf,Lindblom:1998wf,Jaikumar:2008kh}. 
In the general case the damping time due to bulk viscosity reads
\begin{equation}
\tau_B=\frac{2^{3}}{3^{5}5!\pi\alpha^{2}\tilde{J} MR^{2}}\int d^3x \left|\frac{\Delta n}{\bar{n}}\right|^{2}\zeta\left(\left|\frac{\Delta n}{\bar{n}}\right|^{2}\right)
\end{equation}
where $\tilde J$ is a constant 
and the viscosity eq. (\ref{eq:madsen-approximation}) depends on the density fluctuation induced by the r-mode. In the integral we neglect the contribution of the crust of the star since the bulk viscosity has not been computed there, yet.

The temperature dependence of the damping times is shown in fig. \ref{fig:saturation-amplitude} where the bulk viscosity time scale is given at various amplitudes. As can be seen due to the strong increase of the bulk viscosity in the suprathermal regime its damping time decreases strongly with amplitude and eventually undercuts the gravitational time scale. Therefore the enhanced damping can effectively stop the r-mode growth and saturates the amplitude at finite values that are at fixed temperature and frequency determined by the above stability criterium. The result for this "static" saturation amplitude as a function of temperature is shown in fig. \ref{fig:saturation-amplitude} for different frequencies. It features a plateau for temperatures inside the instability region and decreases with frequency. The size of the saturation amplitudes 
should lead to a fast spin down of young stars \cite{Lindblom:1998wf}. Actually, during the r-mode growth the star cools and starts spinning down so that the above amplitudes do not have to be reached. Due to the continuous increase at the right boundary of the instability region the r-mode should automatically saturate at an amplitude that is sufficient for spinning down the star and that could be lower than those of competing mechanisms \cite{Bondarescu:2008qx,Lin:2004wx}.


\begin{theacknowledgments}
We thank Nils Andersson and Andreas Reisenegger for helpful discussions. This research was supported in part by the Offices of Nuclear Physics and High Energy Physics of the
U.S. Department of Energy under contracts
\#DE-FG02-91ER40628,  
\#DE-FG02-05ER41375. 
\end{theacknowledgments}



\bibliographystyle{aipproc}   

\bibliography{cs.bib}

\begin{thebibliography}{13}
\expandafter\ifx\csname natexlab\endcsname\relax\def\natexlab#1{#1}\fi
\providecommand{\enquote}[1]{``#1''}
\expandafter\ifx\csname url\endcsname\relax
  \def\url#1{\texttt{#1}}\fi
\expandafter\ifx\csname urlprefix\endcsname\relax\def\urlprefix{URL }\fi
\providecommand{\eprint}[2][]{\url{#2}}

\bibitem[Papaloizou and Pringle(1978)]{Papaloizou:1978zz}
J.~Papaloizou, and J.~E. Pringle, \emph{Mon. Not. Roy. Astron. Soc.}
  \textbf{182}, 423--442 (1978).

\bibitem[Andersson(1998)]{Andersson:1997xt}
N.~Andersson, \emph{Astrophys. J.} \textbf{502}, 708--713 (1998),
  \eprint{gr-qc/9706075}.

\bibitem[Andersson and Kokkotas(2001)]{Andersson:2000mf}
N.~Andersson, and K.~D. Kokkotas, \emph{Int. J. Mod. Phys.} \textbf{D10},
  381--442 (2001), \eprint{gr-qc/0010102}.

\bibitem[Lindblom et~al.(1999)]{Lindblom:1999yk}
L.~Lindblom, G.~Mendell, and B.~J. Owen, \emph{Phys. Rev.} \textbf{D60}, 064006
  (1999), \eprint{gr-qc/9902052}.

\bibitem[Madsen(1992)]{Madsen:1992sx}
J.~Madsen, \emph{Phys. Rev.} \textbf{D46}, 3290--3295 (1992).

\bibitem[Alford et~al.(2010)]{Alford:2010gw}
M.~G. Alford, S.~Mahmoodifar, and K.~Schwenzer, \emph{J. Phys.} \textbf{G37},
  125202 (2010), \eprint{1005.3769}.

\bibitem[Reisenegger and Bonacic(2003)]{Reisenegger:2003pd}
A.~Reisenegger, and A.~A. Bonacic  (2003), \eprint{astro-ph/0303454}.

\bibitem[Bondarescu et~al.(2009)]{Bondarescu:2008qx}
R.~Bondarescu, S.~A. Teukolsky, and I.~Wasserman, \emph{Phys. Rev.}
  \textbf{D79}, 104003 (2009), \eprint{0809.3448}.

\bibitem[Lin and Suen(2006)]{Lin:2004wx}
L.-M. Lin, and W.-M. Suen, \emph{Mon. Not. Roy. Astron. Soc.} \textbf{370},
  1295--1302 (2006), \eprint{gr-qc/0409037}.

\bibitem[Akmal et~al.(1998)]{Akmal:1998cf}
A.~Akmal, V.~R. Pandharipande, and D.~G. Ravenhall, \emph{Phys. Rev.}
  \textbf{C58}, 1804--1828 (1998), \eprint{nucl-th/9804027}.

\bibitem[Tolman(1939)]{Tolman:1939jz}
R.~C. Tolman, \emph{Phys. Rev.} \textbf{55}, 364--373 (1939).

\bibitem[Lindblom et~al.(1998)]{Lindblom:1998wf}
L.~Lindblom, B.~J. Owen, and S.~M. Morsink, \emph{Phys. Rev. Lett.}
  \textbf{80}, 4843--4846 (1998), \eprint{gr-qc/9803053}.

\bibitem[Jaikumar et~al.(2008)]{Jaikumar:2008kh}
P.~Jaikumar, G.~Rupak, and A.~W. Steiner, \emph{Phys. Rev.} \textbf{D78},
  123007 (2008), \eprint{0806.1005}.

\end{thebibliography}

\IfFileExists{\jobname.bbl}{}
 {\typeout{}
  \typeout{******************************************}
  \typeout{** Please run "bibtex \jobname" to optain}
  \typeout{** the bibliography and then re-run LaTeX}
  \typeout{** twice to fix the references!}
  \typeout{******************************************}
  \typeout{}
 }

\end{document}